\documentclass[12pt]{iopart}
\usepackage{graphicx}% Include figure files

\usepackage{epstopdf}

\usepackage{bm}
\usepackage{dcolumn}% Include figure files
\usepackage{latexsym} 

\newcommand{\D}{\,\mathrm{d}}

\begin{document}

%\draft

\title{Planckian signatures in optical harmonic generation and supercontinuum}
\author{Claudio Conti$^{1,2}$}
\address{$^1$ Department of Physics, University Sapienza, Piazzale Aldo Moro 2, 00185, Rome, IT}
\address{$^2$ Institute for Complex Systems, National Research Council (ISC-CNR), Via dei Taurini 19, 00185, Rome, IT}
\ead{claudio.conti@uniroma1.it}

\vspace{10pt}
\begin{indented}
\item[]\today
\end{indented}

\begin{abstract}
  Many theories about quantum gravity, as string theory, loop quantum gravity, and doubly special relativity, predict the existence of a minimal length scale and
  outline the need to generalize the uncertainty principle.
  This generalized uncertainty principle relies on modified commutation relations that - if applied to the second quantization - imply an excess energy of the electromagnetic quanta with respect to $\hbar \omega$. Here we show that this ``dark energy of the photon'' is amplified during nonlinear optical process.
  Therefore, if one accepts the minimal length scenario, one must expect to observe specific optical frequencies in optical harmonic generation by intense laser fields. Other processes as four-wave mixing and supercontinuum generation may also contain similar spectral features of quantum-gravity.
  Nonlinear optics may hence be helpful to falsify some of the most investigated approaches to the unification of quantum mechanics and general relativity.
\end{abstract}


\maketitle
\section{Introduction}
The search for observable signatures of the new physics at the Planck scale is a route we must follow to test the many ideas for the unification of quantum mechanics and general relativity.~\cite{MukhanovBook2007, BookRovelli2015, ThiemannBook2007, BookKiefer2012}
As reviewed in \cite{Amelino-Camelia2013}, there are many proposed tests of quantum-gravity phenomenology, including quantum-optics \cite{Gambini99, Alfaro02, Rideout2012}.

Despite these investigations, it is hard to believe to the possibility of probing Planckian physics in the laboratory. The use of intense laser fields and the corresponding impressive technological developments of recent years may support our imagination. This possibility was discussed in particular by Magueijo.~\cite{Magueijo2006}
At the moment, most of the emphasis is in laser-driven particle acceleration, laser induced particle-antiparticle generation, and related quantum-field processes. This activity gains momentum by the realization of novel extreme light infrastructures.~\cite{Kuehn2017}

Here we consider a different perspective, and try to show that a particular
effect predicted by many different quantum-gravity theories may be falsified by experiments in nonlinear optics. \\ The considered effect is related to the existence of a minimal length scale, which has been predicted by string-theory, loop quantum gravity, doubly-special relativity, polymer quantization, black hole physics and related investigations.\cite{Hossenfelder2013}

The minimal length scenario has important consequences. In order to include this scenario in our fundamental models, we need to change the Heisenberg uncertainty principle. In simple terms, the existence of a minimal length is due to granularity of the spacetime quantified by the Planck scale $\ell_P$ (or other unknown length scale). We cannot localize particles at a length scale smaller than $\ell_P$. This fact - as explored in the vast related literature - is at odds with the standard uncertainty principle, which does not predict any minimal value for the position uncertainty $\Delta X$ (we neglect here relativistic effects \cite{LandauBookQED}).

The most studied generalized uncertainty principle (GUP) reads as
\begin{equation}
  \Delta X \Delta P_X \ge \frac{\hbar^2}{2}\left(1+\beta_X \Delta P_X^2\right)
    \label{GUP}
  \end{equation}
  with $\hat P_X$ the momentum, and $\beta_X$ a unknown parameter. $\beta_X$ fixes the minimal position uncertainty $\Delta X_{min}=\hbar\sqrt{\beta_X}$.
  Since the original proposals,~\cite{YONEYA1989, Amati1989, Maggiore1993, Garay1995, Kempf95, Scardigli99} the possibility of a generalized uncertainty principle has attracted a lot of attention, ranging from theoretical works to proposed experimental tests.~\cite{Chang02, Das08,Pedram12,  Das08, Battisti09, Pikovski12, Marin2015, Marcucci2018, Khodadi2018, Khodadi2018a, Scardigli2017, Scardigli2018}

  A key-strategy to account for Eq.~(\ref{GUP}) by ``minimal changes'' to the standard quantum mechanics is to consider modified commutators. Standard quantum mechanics with the generalized commutator
  \begin{equation}
    [\hat X, \hat P_X]=\imath \hbar (1+\beta_X \hat P_X^2)
    \label{MUP}
    \end{equation}
    readily implies Eq.~(\ref{GUP}).~\cite{Kempf95}
    These generalized quantum mechanics have been largely studied in recent years, looking for observable phenomena in the laboratory, or at an astrophysical scale. One intriguing outcome of modified commutation relations as (\ref{MUP}) is the shift of the energy eigenvalues of elementary quantum mechanics models as, specifically, the harmonic oscillator.\cite{Brau1999, Chang02, Das09}

    Here we follow Refs.~\cite{Kempf95,PEDRAM2013}, and adopt modified uncertainty principles in the quantization of the electromagnetic (EM) field. 
    The energy of the photons is shifted by the quantum-gravity terms in the modified commutation relations. As other authors pointed out,\cite{KempfMangano97, Chang02} this shift changes the black-body spectrum and other equilibrium phenomena.
    We investigate by simple arguments the effect of the energy shift of light quanta in optical harmonic generation. We consider the scenario in which this excess (``dark'') energy of the photon is emitted as novel photons.
    If this scenario holds true, one should observe a new kind of EM radiation whose spectral content changes with the laser pulse energy.
    This looks to be an observable phenomenon, which can perhaps falsify the minimal length theory, or set quantitative bounds for the modifications to the conventional field quantization.
    
    Various laboratories are developing approaches for optical ultra-broadband generation by novel nonlinear devices and materials 
with the observation of complex and unexpected spectral content.\cite{Silva2012,Belli2015,Bang2016} The possibility that these new frequencies are signatures of the new physics at the Planck scale is certainly exciting and worth to be investigated.   
\section{Modified Quantization of the Electromagnetic Field}
We start from the standard quantization procedure for the electromagnetic (EM) field.~\cite{Dirac1927,Loudon2000}
In a cavity, the energy of the classical electrical field with angular frequency $\omega_{\mathbf k}$ is written as
\begin{equation}
  \mathcal{E}=\frac{1}{2}\sum_{\mathbf k} P_{\mathbf k}^2+\omega_{\mathbf k}^2 Q_{\mathbf k}^2,  
\end{equation}
where $P_{\mathbf k}$ and $Q_{\mathbf k}$ are the field quadratures, and ${\mathbf k}$ is the mode index.
In the following, we consider a single mode and omit the index ${\mathbf k}$:
\begin{equation}
  \mathcal{E}=\frac{1}{2}\left( P^2+\omega^2 Q^2  \right).
  \label{EHO}
\end{equation}

As reported in quantum optics textbooks (see, e.g.,~\cite{Loudon2000}), following the original Dirac strategy \cite{Dirac1927}, one recognizes in Eq.(\ref{EHO}) a formal equivalence with the harmonic oscillator, and quantizes the field by
converting the classical quantities $P$ and $Q$ in operators with the commutation relation $[\hat Q,\hat P]=i\hbar$.
Correspondingly, the EM quantized energy is
\begin{equation}
  E_n=\hbar\omega \left(n+\frac{1}{2}\right).
  \end{equation}
  For each photon with frequency $\omega$, there is a quantum of energy $\hbar \omega$.

  We hence follows the same argument, and identify in Eq.~(\ref{EHO}) a quantum harmonic oscillator. However, with the prescription of the mentioned quantum gravity theories, we adopt a modified commutator for $\hat P$ and $\hat Q$, namely
  \begin{equation}
    [\hat Q, \hat P]=\imath \hbar \left(1+\beta \hat P^2\right).
    \label{GCOMM}
    \end{equation}
    In the GUP literature, the quantum harmonic oscillator has been revised and largely studied. Pedram~\cite{Pedram2012} has shown that one can write $\hat Q$ and $\hat P$ in terms of the standard $\hat q$ and $\hat p$, with $[\hat q, \hat p]=\imath \hbar$, as
    \begin{equation}
      \begin{array}{l}
        \displaystyle\hat Q=\hat q\\
        \displaystyle\hat P=\frac{\tan(\sqrt{\beta}\hat p)}{\sqrt{\beta}},
      \end{array}
      \label{Qq}
      \end{equation}
    and determine analytically the energy of the modified harmonic oscillator (see appendix).
\section{The ``dark energy'' of the photon}
In the modified quantum mechanics with the generalized commutator in Eq.~(\ref{GCOMM}), the eigenstates and eigenvalues of the energy are also modified.
We follow refs.~\cite{Pedram2012,PEDRAM2013} for the harmonic oscillator, the energy eigenvalues are \cite{Kempf95}
\begin{equation}
  E_n(\omega)=\hbar\omega\left(n+\frac{1}{2}\right)\left(\sqrt{1+\frac{\beta^2 \hbar^2\omega^2}{4}}+\frac{\beta\hbar\omega}{2}\right)+\frac{\beta\hbar^2\omega^2}{2}n^2.
  \label{GHOE}
\end{equation}
For $\beta=0$, Eq.~(\ref{GHOE}) gives the well known $E_n=(n+1/2)\hbar\omega$.
 At the lowest order in $\beta$, we have
\begin{equation}
  E_n(\omega)\simeq\hbar\omega\left(n+\frac{1}{2}\right)+\beta\frac{\hbar^2\omega^2}{4}\left(2n^2+2n+1\right).
  \label{GHOE1}
\end{equation}
Eq.~(\ref{GHOE1}) shows the modified dispersion expected to be valid at the Planck scale, following the recipe of quantum-gravity theories, as string-theory.~\cite{Witten1996} Eq.~(\ref{GHOE1}) also shows that one can retain the standard expression $E_n=n \hbar\omega_n$, if the angular frequency is assumed to be dependent on the number of photons, with
\begin{equation}
  \omega_n\equiv \frac{E_n(\omega)}{n \hbar}.
\end{equation}
For $n>>1/2$ we have
\begin{equation}
  \omega_n\simeq \omega\left( 1+n \beta \frac{\hbar\omega}{2}\right),
  \label{NLOMEGA}
\end{equation}
which signals a blue-shift of the photon energy when the number of photons grows.
This result may be related to similar phenomena within doubly-special relativity (see, for example, \cite{Amelia17} and references therein).  

  From Eq.~(\ref{GHOE}), we have $E_n(\omega)>n\hbar\omega$.
  This implies that a single photon at frequency $\omega$ has an {\it excess energy} with respect to the standard quantum mechanics.
Apparently, this excess energy does not correspond to any detectable electromagnetic color, we hence refer to it as the ``dark-energy of the photon''.
  For the single photon, we have
\begin{equation}
  \delta_1 E(\omega)=E_1(\omega)-\frac{3}{2}\hbar\omega\simeq\frac{5}{4}\beta\hbar^2\omega^2,
  \label{dark1}
\end{equation}
and, for $n$ photons with $n>>1$,
\begin{equation}
  \delta_n E(\omega)=E_n(\omega)-\left(n+\frac{1}{2}\right)\hbar\omega\simeq \beta n^2 \frac{\hbar^2 \omega^2}{2}.
  \label{dark2}
  \end{equation}
  In general, this dark energy is very small and - at equilibrium - is seemingly impossible to observe. We may argue if $\delta E_n$ has some observable signature.
  
One can consider many different effects as, for example, cosmological dark-energy, black-body radiation, quantum noise, spontaneous and stimulated emission processes. $\delta E_n$ is very small, and lab-top experiments aimed to test this excess energy are difficult to imagine and realize.
Looking at Eq.(\ref{dark2}), it is natural to consider high-field effects, as nonlinear electromagnetic processes, because $\delta E_n$ scales with the square of the number of photons $n$.

In the next section, we consider the second-harmonic generation (SHG) of optical radiation. This process was also previously considered for lab-tests of the modified dispersion relation predicted by doubly special relativity.\cite{Amelino-Camelia2013}
\subsection{The dark energy of the photon and the dark energy of the universe}
Following the standard model of cosmology, about $70\%$ of the total energy of the universe is unknown and has very low density. The density of the dark energy is estimated to be $10^{-27}$~kg~m$^{-3}$.~\cite{WeinbergCosmology}

We want to compare the density of the dark energy with the density of $\delta E_1$. The largest contribution to the photons in the universe comes from the cosmic microwave background (CMB), with $\omega_{CMB}\simeq 10^{12}$~rad/s and a total number of photons $n_{CMB}=10^{89}$. By using Eq.~({\ref{dark1}}), we have that the ``dark energy of the CMB'' is
\begin{equation}
  \delta E'_{CMB}\equiv n_{CMB}\delta E_1(\omega_{CMB})\simeq n_{CMB}\beta(\hbar\omega_{CMB})^2.
  \label{deltaprime}
  \end{equation}
In (\ref{deltaprime}), we assume that the CMB photons belong to different modes. For $\beta=1/\mathcal{E}_P$, with $\mathcal{E}_P$ the Planck energy, taking the volume of the universe as $V_u=10^{80}$m$^3$, we have density $\rho'_{CMB}=\delta E'_{CMB}/(c_0^2 V_u)=10^{-62}$~kg~m$^{-3}$, with $c_0$ the vacuum light velocity. $\rho'_{CMB}$ is much smaller than the density of the dark energy.

  Recent models of dark matter accounted for the possibility of large scale Bose-Einstein condensates (see, for example,~\cite{Paredes2016} and reference therein). If we hence take all the CMB photons condensed in the same mode (Eq.~(\ref{dark2}))
  \begin{equation}
  \delta E''_{CMB}\equiv \delta E_{n_{CMB}}(\omega_{CMB})\simeq n^2_{CMB}\beta(\hbar\omega_{CMB})^2 ,
  \end{equation}
  which gives the too large $\rho''_{CMB}=10^{27}$~kg~m$^{-3}$.

  These calculations are done with the arbitrarily chosen $\beta=1/\mathcal{E}_P$; on the contrary, one can use the $\beta$ parameter to fit the density of dark-energy. For example, in the case of the condensate of dark matter, when choosing $\beta=10^{-64}$J$^{-1}$,  the dark energy of the CMB photons $\delta E''_{CMB}$ has the same density of the dark energy in the universe. 
\section{Excess energy in second harmonic generation}
In optically nonlinear media, intense electromagnetic fields may combine
and generate novel frequencies.~\cite{BloembergenBook} The simplest process is the second-harmonic generation: $2$ photons with frequency $\omega$ generate a single photon at frequency $2\omega$. In the standard quantum mechanics, such a process readily conserves energy, that is, the energy of the generated photon $2\hbar\omega$ is the sum of the energies  $\hbar\omega$ of the $2$ original photons.

This simple scenario is however more complicated in the modified quantum mechanics here considered. As detailed in the following,
excess energy is to be expected.
\subsection{``Type II'' second-harmonic generation}
We start considering the case in which two photons at the fundamental frequency (FF) $\omega$ have different polarization,
and hence belong to distinguishable modes. The two photons combine to generate a second-harmonic photon (SH).
In the literature about nonlinear optics, this case is conventionally named ``type II'' SHG.
The energy of the two photons is hence $2 E_1(\omega)$, and it turns out to be different from the SH photon energy $E_1(2\omega)$.
Specifically, we have
\begin{equation}
 E_1(2\omega)> 2E_1(\omega).
\end{equation}
Therefore, we have the paradox that {\it optical second harmonic generation does not conserve energy}.

If we still assume that the GUP is valid, the solution for this paradox may be that the excess energy
\begin{equation}
  \delta E_1^{SHG-II}=E_1(2\omega)-2E_1(\omega)
  \end{equation}
  is lost in other degrees of freedom as, e.g., thermal or acoustic energy of the crystal adopted for the SHG.
  
  This possibility is very difficult to explore in experiments. For example, one can imagine to make a SHG experiment at very low temperature and look for excess heat. However, even at very low temperature, it may be impossible to distinguish this heat from the one due to the linear absorption of photons, which is always present.

  Here we consider the possibility that the excess energy  $\delta E^{SHG-II}$ is still available as electromagnetic degrees of freedoms.
  This corresponds to emission of frequency-shifted excess photons during SHG.

  We consider the case in which $2n$ photons at frequency $\omega$ are converted to $n$ photons at $2\omega$. The excess energy for $n$ second-harmonic photons reads
\begin{equation}
  \delta E_n^{SHG-II}=E_{n}(2\omega)-2E_{n}(\omega).
  \label{dSHG}
\end{equation}
For $n>>1$, at lowest order in $\beta$, we have
\begin{equation}
  \delta E_n^{SHG-II}\simeq \beta (\hbar \omega)^2 n^2
  \label{dSHGb}
\end{equation}
Eq.~(\ref{dSHGb}) shows that the excess energy scales quadratically with the number of photons, which is a very convenient situation as $\beta$ can be very small. 

\subsection{Estimate of the emission spectrum}
The theories for quantum gravity do not predict the value for $\beta$. However, this value is expected to be related to the Planck scales. In particular, as the $\beta$ here considered has the dimension of the inverse of energy, we consider the Planck energy $\mathcal{E}_P\cong 10^9$J and assume, as a representative case, $\beta\simeq 1/\mathcal{E}_P$. SHG and related experiments may be adopted to set limits to $\beta$.

Our hypothesis is that, if the GUP scenario is correct, the excess energy is converted in emitted radiation. We can estimate the wavelength of the emitted radiation by assuming that the excess energy $\delta E_n^{SHG-II}$ is found in $N$ photons in modes at angular frequency $\Omega$. We can estimate $\Omega$ by the equation
\begin{equation}
  E_N(\Omega)=\delta E_n^{SHG-II}(\omega).
  \end{equation}
  Considering Eq.(\ref{dSHGb}), and neglecting the $\beta$ correction for the energy of the  photons at $\Omega$, we have
\begin{equation}
  N\hbar \Omega =\beta (\hbar \omega)^2 n^2 .
  \label{Omega1}
  \end{equation}
  Eq.~(\ref{Omega1}) predicts the energy of the generated modes and shows that, even if $\beta$ is very small, a sufficiently large number of pumping photons
may lead to an observable emission due to the Planck scale effect.

\begin{figure}
\centering
    \includegraphics{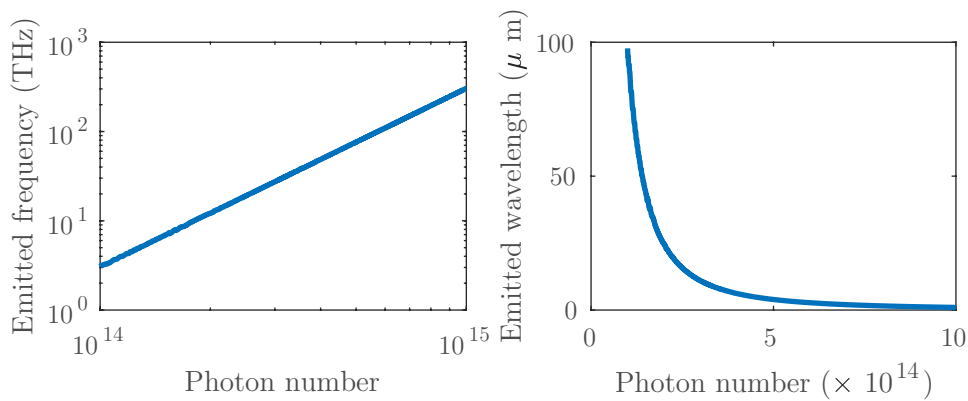}
\caption{
  Estimated emission frequency $\Omega/2\pi$ (left panel) and wavelength (right panel) for $N=100$ output photons versus the number of photons in the pump pulse for type-II second-harmonic generation.}
\label{figure1}
\end{figure}

Assuming that a observable amount of photons at $\Omega$ is generated (we take $N=100$), figure \ref{figure1} shows the emitted frequency when varying the photon number $n$ and taking $\omega=2\pi c_0/\lambda$ with $\lambda=1\mu$m and $\beta=1/\mathcal{E}_P$.

It is remarkable that, with the small value chosen for $\beta$, one obtains observable frequencies in the optical domain at a moderate pump photon number $n$.
For example, $n=10^{15}$ photons are contained in a laser pulse with energy of the order of $100\mu$J, which is routinely employed in ultra-fast nonlinear optics.

The experimental signature is given by the shift of the generated frequencies $\Omega/2\pi$ with the pulse energy (or photon number $n$). The slope of the shift gives a estimate for the unknown $\beta$ as in Eq.~(\ref{Omega1}).
Notably the emitted frequency may fall in the UV region, or lower wavelengths, when increasing $n$ (for the considered value of $\beta$).
\subsection{ ``Type I'' second harmonic generation}
In this case, we assume that FF photons belonging to the same mode are converted to the second harmonic.
In nonlinear optics, this is commonly indicated as ``type 0,'' or ``type I,'' SHG.
In this case, we have $2n$ photons with angular frequency $\omega$ in a mode with energy $E_{2n}(\omega)$.

By using Eq.~(\ref{GHOE}), we have
\begin{equation}
  \delta E_n^{SHG-I}=E_{n}(2\omega)-E_{2n}(\omega).
\end{equation}
For $n>>1$, at the lowest order in $\beta$, we have
\begin{equation}
  \delta E_n^{SHG-I}\simeq \beta (\hbar \omega)^2 n,
\end{equation}
which scales linearly with the number of photons $n$. Hence, when $n\simeq 10^{15}$, $\delta E_n^{SHG-I}$ is several order of magnitudes smaller than   $\delta E_n^{SHG-II}$ considered above.
\section{Third order nonlinearity and Planck-scale Kerr effect}
At this stage, we identify a further paradox: we see that a nonlinear-optical process at the second order, which commonly involves $3$ photons, resembles a third order process and involves $4$ photons. Indeed, we are claiming that we generate photons at frequency $\Omega$, during second harmonic generation with couples of photons at $\omega$ and generated photons at $2\omega$. 
The paradox is solved by the fact that the generalized commutators actually induce a third-order nonlinearity in the field evolution and, hence, new frequencies can be generated in addition to $\omega$ and $2\omega$.

We introduce the wave-vector as $k_n=\omega_n/ c$ and, defining the refractive index $n_R$, we set $k_n=\omega_n/c=\omega n_R/c $, which gives the refractive index of vacuum as
\begin{equation}
  n_R=1+\frac{1}{2}\beta \hbar\omega n.
  \label{n2GUP}
  \end{equation}
Eq.~(\ref{n2GUP}) resembles the well known third-order optical Kerr effect,~\cite{BoydBook}, i.e., a energy dependent refractive index. Eq.~(\ref{n2GUP}) also reveals the vacuum polarizability arising from the existence of a minimal length. Vacuum Kerr effect due to high-energy effects (as particle generation) was considered by many authors in the past (see, e.g., \cite{Battesti2013}).

In a different perspective, we can consider the Hamiltonian in Eq.~(\ref{Qq}) in terms of the conventional $\hat q$ and $\hat p$:
\begin{equation}
  \hat H=\frac{1}{2}\left(\hat P^2+\omega^2 \hat Q^2\right)=\frac{1}{2}
  \left\{\left[\frac{\tan(\sqrt{\beta}\hat p)}{\sqrt\beta}\right]^2+\omega^2 \hat q^2 \right\}.
  \end{equation}
  At the lowest order in $\beta$, we have 
  \begin{equation}
    H=\frac{1}{2}\left(\hat p^2 +\omega^2 \hat q^2\right)+\frac{\beta}{3}\hat{p}^4.
    \end{equation}
    We introduce the standard ladder operators $\hat a$ and $\hat a^{\dagger}$
    \begin{equation}
      \begin{array}{l}
        \hat q=\sqrt{\frac{\hbar}{2\omega}}
        \left(\hat a^{\dagger}+\hat a \right)\\
        \hat p=\imath \sqrt{\frac{\hbar\omega}{2}}
        \left(\hat a^{\dagger}-\hat a \right) ,
        \end{array}
      \end{equation}
     and we have an Hamiltonian with third order nonlinearity :
      \begin{equation}
        \hat H=\frac{\hbar\omega}{2}\left(\hat a^{\dagger}\hat a +\frac{1}{2}\right)+\beta H_1.
        \end{equation}
        Indeed, after normal ordering,
        \begin{equation}
H_1=\frac{(\hbar \omega)^2}{12} \left[
          \hat a^4- 4 \hat a^{\dagger} a^3+ 6 (\hat a^{\dagger})^2 (\hat a)^2
          - 4 (\hat a^{\dagger})^3 a+(\hat a^{\dagger})^4\right].
        \label{nlH}
\end{equation}
Eq.(\ref{nlH}) shows that modified uncertainty relations introduce higher order four-wave mixing terms \cite{Pedram2010}; therefore, we may expect the generation of novel photons.
  \section{Link with supercontinuum generation}
The previous arguments are obviously not limited to the process of SHG, but they readily apply to more complex nonlinear process as, e.g., supercontinuum, high-harmonic generation, etc.
To show the way these arguments may be extended, let us consider a four-wave-mixing process - the onset of supercontinuum - in which $3$ angular frequencies $\omega_1$, $\omega_2$ and $\omega_3$ combine to generate a fourth wave at angular frequency $\omega_4$. In the standard formulation ($\beta=0$), energy conservation implies $\omega_4=\omega_1+\omega_2+\omega_3$. However, in our framework, the excess energy reads
\begin{equation}
  \delta E_1^{4WM}(\omega_4)=E_1(\omega_4)-E_1(\omega_3)-E_1(\omega_2)-E_1(\omega_1),
\end{equation}
which can be used to generalize the arguments above to other nonlinear processes.

\section{Conclusions}
We have assumed the validity of the arguments concerning the existence of a minimal length scale coming from string theories, loop quantum gravity, doubly special relativity, and other theories attempting to unify general relativity and quantum mechanics. As discussed by several authors, these arguments imply a generalization of the uncertainty principle, and hence of the standard commutation relations. The way the commutation relation has to be changed (Eq.~(\ref{GCOMM})) follows the most investigated formulation.

We hence postulated that these generalized commutators must be also valid for the electromagnetic field quadratures and - in a more general perspective - to any other classical field quantization.

This approach leads to a perturbation to the energy of the quanta of the electromagnetic field, according to Eq.$(\ref{GHOE})$. This perturbation is extremely small, corresponds to an excess ``dark energy'' with respect to the commonly accepted value $\hbar\omega$, and links to the discussion about dark energy and related cosmological phenomena.

We tried to identify other phenomena that can be tested in the laboratory, and we considered nonlinear optics. By simple arguments, it turn outs that - if Eq.(\ref{GHOE}) is valid - common frequency mixing processes do not conserve energy. This conclusion may eventually rule out the validity of a generalized uncertainty principle.

But if we still assume that Eq.(\ref{GHOE}) is valid, we can quantify what happens if the excess energy in optical frequency conversion is emitted as photons. An encouraging outcome is the fact that
the frequency of the generated modes grows with the number $n$ of the pumping photons. This circumstance gives a direct smoking gun of the ``Planckian emission'' and may be searched in the experiments. In general, the very large number of photons attainable with modern laser technology may allow to unveil the very small effects weighted by the unknown constant $\beta$.

Highly nonlinear optical processes, as supercontinuum generation, driven by modern devices, like micro-structured optical fibers, may furnish novel roads for looking at exotic - but fundamental - phenomena, as the effects of quantum-gravity in the electromagnetic field propagation. There are open questions in recent experiments concerning ultra-wide band generation with ultra-short pulses. Looking for out-of-shelf explanations may be an interesting adventure.
\section{Acknowledgments}
This publication was made possible through the support
of a grant from the John Templeton Foundation (grant
number 58277). The opinions expressed in this publication are those of the author and do not necessarily reflect the views of the John Templeton Foundation.
\appendix
\section*{Appendix : The modified harmonic oscillator}
\setcounter{section}{1}
Various authors in the literature about the generalized uncertainty principle investigated the modifications to the quantum harmonic oscillator since the early papers.~\cite{Kempf95,Brau1999,Chang02}
Pedram~\cite{Pedram2012} reported on a elegant formulation of the problem by expressing the generalized quadratures $\hat Q$ and $\hat P$ in terms of the standard ones $\hat q$ and $\hat p$ by Eq.~(\ref{Qq}).

If one considers the Hamiltonian
\begin{equation}
  \hat H=\frac{1}{2}\left(\hat P^2+\omega^2 \hat Q^2 \right)
  \end{equation}
  with the modified commutation relation
  \begin{equation}
    \left[\hat P, \hat Q\right]=\imath \hbar \left(1+\beta \hat P^2\right),
  \end{equation}
  it turns out that the eigenproblem is exactly solvable by letting
      \begin{equation}
      \begin{array}{l}
        \displaystyle\hat Q=\hat q\\
        \displaystyle\hat P=\frac{\tan(\sqrt{\beta}\hat p)}{\sqrt{\beta}},
      \end{array}
      \label{Qq1}
      \end{equation}
      with $\left[\hat p, \hat q\right]=\imath \hbar$.
        In the momentum representation, the stationary Schr\"odinger equation reads
        \begin{equation}
          -\frac{\hbar^2\omega^2}{2}\frac{\D^2 \phi}{\D p^2}+\frac{1}{2}\frac{tan^2\left(\sqrt{\beta p}\right)}{\beta}\phi=E \phi.
          \label{pHO}
          \end{equation}
          As detailed in \cite{Pedram2012}, equation (\ref{pHO}) is analytically solved in terms of the Gauss hypergeometric functions \cite{Taseli2003}, and the eigenvalues are given by Eq.(\ref{GHOE}).

\section*{References}
%\bibliographystyle{iopart-num}
%\bibliography{../../../bibtex/GIGAbib}

\providecommand{\newblock}{}

\end{document}